\documentclass[12pt]{article}
\usepackage{epsfig}
\usepackage{psfrag}
\usepackage{a4}

\newcommand{\be}{\begin{equation}}
\newcommand{\ee}{\end{equation}}

\begin{document}
\title{Hedged Monte-Carlo: low variance derivative pricing with objective probabilities}

\author{Marc Potters$^{\dagger}$, Jean-Philippe
Bouchaud$^{\dagger,*}$ and Dragan Sestovic$^{\dagger}$}
\date{{\small $^\dagger$ Science \& Finance, Capital Fund Management,\\
109-111 rue Victor Hugo, 92532
Levallois {\sc cedex}, FRANCE\\ http://www.science-finance.fr\\
$^*$ Service de Physique de l'\'Etat Condens\'e,
 Centre d'\'etudes de Saclay,\\
Orme des Merisiers,
91191 Gif-sur-Yvette {\sc cedex}, FRANCE\\}
\vspace{0.5cm}
August 9, 2000}
\maketitle

\begin{abstract}
We propose a new `hedged' Monte-Carlo ({\sc hmc}) method to price financial derivatives, which
allows to determine simultaneously the optimal hedge. The inclusion of the optimal hedging
strategy allows one to reduce the financial risk associated with option trading, and 
for the very same reason reduces considerably the variance of our {\sc hmc} scheme as compared to
previous methods. The explicit accounting of the hedging cost naturally converts 
the objective probability into the `risk-neutral' one. This allows a consistent use 
of purely historical time series to price derivatives and obtain their residual risk.
The method can be used to price a large class of exotic options, including those with
path dependent and early exercise features.
\end{abstract}

\section{Introduction}

The Black and Scholes model for options has two truly remarkable
properties: (i) one can find a `perfect' hedging strategy that eliminates risk entirely, 
and (ii) the option price does not depend at all on the average return of the
underlying asset \cite{BSc,Hull,Wilmott}. This last property shows that the option 
price is not simply the (discounted) average of the future pay-off over 
the objective (or historical) probability distribution, as one would have naively
expected. This is even more striking in the case of the Cox-Ross-Rubinstein 
binomial model \cite{CRR,Hull} where the pricing measure is
completely unrelated to the actual distribution of returns. These two models have
deeply influenced financial mathematics, and have lead to the development of a rather abstract 
and general framework for derivative pricing. Within this framework, the fundamental result 
is that the requirement of absence of arbitrage opportunities is equivalent to the existence of a `risk-neutral
probability measure' ({\it a priori} distinct from the objective one), such that the price 
of a derivative is indeed its (discounted) average pay-off, but where the average is performed 
over the risk-neutral distribution rather than over the objective distribution \cite{Pliska,Baxter}. 
It is thus a common belief that the knowledge of the `true' 
probability distribution of returns is a useless information to price options.
The credence is rather that the relevant risk neutral distribution is somehow `guessed'
by the market. A possibility is to reconstruct this risk neutral distribution from 
quoted option prices, and to use this information to price other
instruments consistently (for example exotic options) \cite{Hull,Dupire}. 
In many cases, however, exact analytical expressions for these exotic options
are not available and numerical methods must be used. A very versatile method, which
allows to price complicated path dependent options, is the Monte-Carlo method,
where paths are generated with a weight consistent with the risk-neutral distribution.
Efficient numerical techniques have been developed, in particular in \cite{Schwartz}, to
implement this `risk neutral Monte-Carlo' ({\sc rnmc}).

However, in most models of stock fluctuations, except for very special
cases, risk in option trading cannot be eliminated, and strict
arbitrage opportunities do not exist, whatever the price of the
option. That risk cannot be eliminated is furthermore the fundamental
reason for the very existence of option markets. It would thus be more
satisfactory to have a theory of options where the objective
stochastic process followed by the underlying asset was used to
compute the option price, the hedging strategy, and {\it the residual
risk}. The latter is clearly important to estimate for risk control
purposes.  A natural framework for this is the risk minimization
approach developed by several authors \cite{Schw,Schal,BS,BP,AS},
where the optimal trading strategy is determined such that the chosen
measure of risk is minimized (for example the variance of the wealth
balance, although other choices are possible \cite{BP,SB}) . The
`theoretical' price is then obtained using a fair game
argument. Interestingly, this framework allows one to recover exactly
the Black-Scholes results when the objective probabilities are
log-normal, and when the continuous time limit is taken (this is shown
in details in \cite{BP}). In particular, the average trend indeed
completely disappears from the price and hedge.

The aim of this paper is to present a very general Monte-Carlo scheme based on this
approach, that we call `hedged Monte-Carlo' ({\sc hmc}).
The method, which has been inspired in part by the least square method ({\sc lsm}) of Longstaff and
Schwartz \cite{Schwartz}, shares with it the property that it can price a wide
variety of exotic options including those with path-dependent or early
exercise features. On top of that, the {\sc hmc}
has at least four major advantages over {\sc rnmc}:
\begin{itemize}
\item The most important one is considerable variance reduction. This is related to the fact
that the financial risk arising from the imperfect replication of the option by the
hedging strategy is directly related to the variance of the Monte-Carlo simulation. 
When minimizing the former by choosing the optimal strategy, we automatically reduce the latter.
The standard deviation of our results are typically five to ten times smaller than with {\sc rnmc},
which means that for the same level of precision, the number of trajectories needed in the 
Monte-Carlo is up to a hundred times smaller.

\item The {\sc hmc} method provides not only a numerical estimate of
the price of the derivative, but also of the optimal hedge (which may
be different from the Black-Scholes $\Delta$-hedge for non Gaussian
statistics) and of the residual risk.

\item The method does not rely on the notion of risk neutral measure
and can be used with any model of the true dynamics of the underlying (even
very complex), in
particular those for which the risk neutral measure is unknown and/or not
uniquely defined.

\item Last, but not least, the {\sc hmc} method allows one to use
purely historical data to price derivatives, short-circuiting the
modeling of the underlying asset fluctuations.  These fluctuations are
known to be of a rather complex statistical nature, with fat-tailed
distributions, long-range volatility correlations, negative
return-volatility correlations, etc. \cite{Review,MS,BP,Muzy}. Within
the {\sc hmc} method, one can directly use the historical time series
of the asset to generate the paths. The fact that a rather small
number of paths is needed to reach good accuracy means that the length
of the historical time series need not be very large.

\end{itemize}

\section{Basic principles of the method}

Option pricing always requires to work backwards in time. This is because the option
price is exactly known at maturity, where it is equal to the pay-off. As with other schemes,
we determine the option price by working step by step for maturity $t=N \tau$ to the present time
$t=0$, the unit of time $\tau$ being, for example, one day. The price of the underlying asset at
time $k\tau$ is denoted as $x_k$ and the price of the derivative is ${\cal C}_k$. We assume for
simplicity that ${\cal C}_k$ only depends on $x_k$ (and of course on $k$). However, the method 
can be generalized to account for a dependence of ${\cal C}_k$ on the volatility, interest rate, etc.,
or to price multi-dimensional options (such as interest rate derivatives). We therefore also
introduce the hedge $\phi_k(x_k)$, which is the number of underlying asset in the portfolio
at time $k$ when the price is equal to $x_k$. Within a quadratic measure of risk, 
the price and the hedging strategy at time $k$ is such that the variance of the wealth 
change $\delta W_k$ between $k$ and $k+1$ is minimized. More precisely, we define the local
`risk' ${\cal R}_k$ as:
\be
{\cal R}_k = \langle \left({\cal C}_{k+1}(x_{k+1})-{\cal C}_k(x_{k})+\phi_k(x_k)[x_k-x_{k+1}]\right)^2
\rangle_o,\label{fond}
\ee
where $\langle ... \rangle_o$ means that we average over the objective probability measure (and
{\it not} the risk-neutral one!). As 
shown in \cite{BP}, the functional minimization of ${\cal R}_k$ with respect to both 
${\cal C}_k(x_k)$ and $\phi_k(x_k)$ gives equations that allow one determine the price and hedge,
provided ${\cal C}_{k+1}$ is known. Note that we have not included interest rate effects in 
Eq.~(\ref{fond}). When the interest rate $r$ is non zero, one should consider the following 
modified equation:
\be{\cal R}_k = \left\langle \left(e^{-\rho}{\cal C}_{k+1}(x_{k+1})-{\cal C}_k(x_{k})
+\phi_k(x_k)[x_k-e^{-\rho}x_{k+1}]\right)^2
\right\rangle_o,\label{fondwrates}
\ee
where 
$\rho=r\tau$ is the interest rate over an elementary time step $\tau$.

In order to implement this numerically, we parallel the {\sc lsm} method of Long\-staff and Schwartz, 
developed within a risk-neutral approach \cite{Schwartz}.
We generate a set of $N_{\rm MC}$ Monte-Carlo trajectories $x^\ell_k$, where $k$ is the time index and $\ell$
the trajectory index. We decompose the functions ${\cal C}_k$ 
and $\phi_k$
over a set of $M$ appropriate basis functions $C_a(x)$ and $F_a(x)$:\footnote{For numerical
purposes, these basis functions have been chosen to be piecewise linear for $F_a$ and piecewise
quadratic for $C_a$, with adaptive breakpoints.}
\be
{\cal C}_k(x) = \sum_{a=1}^M \gamma_a^k C_a(x) \qquad 
{\phi}_k(x) = \sum_{a=1}^M \varphi_a^k F_a(x).
\ee
In other words, we solve the minimization problem with the {\it variational space} spanned by
the functions $C_a(x)$ and $F_a(x)$. This leads to a major simplification since now we have
a linear optimization problem in terms of the coefficients $\gamma_a^k, \varphi_a^k$, for which
efficient numerical algorithms are available \cite{Numrec}. These coefficients must be such that:
\be\label{EQ4}
\sum_{\ell=1}^{N_{\rm MC}}
\left(e^{-\rho}{\cal C}_{k+1}(x^\ell_{k+1})-\sum_{a=1}^M \gamma_a^k C_a(x^\ell_{k})+\sum_{a=1}^M \varphi_a^k 
F_a(x^\ell_k)[x^\ell_k-e^{-\rho}x^\ell_{k+1}]\right)^2 
\ee
is minimized. Those $N$ minimization problems (one for each $k=0,...,N-1$) 
are solved
working backwards in time with ${\cal C}_N(x)$ the known final pay-off function.

Although in general the optimal strategy is not equal to the Black-Scholes $\Delta$-hedge, the difference
between the two is often small, and only leads to a second order increase of the risk \cite{BP}. 
Therefore, one can choose to work within a smaller variational space and impose that:
\be
\varphi_a^k \equiv \gamma_a^k \qquad F_a(x)\equiv \frac{dC_a(x)}{dx}.
\ee
This will lead to exact results only for Gaussian processes, but reduces the computation cost by a
factor two.

\section{Numerical results for the Black-Scholes model}

We have first checked our {\sc hmc} scheme when the paths are
realizations of a (discretized) geometric random walk. We have priced
an at the money three month European option, on an asset with $30 \%$
annualized volatility and a drift equal to the risk-free rate which we
set to 5 \% per annum. The number of time intervals $N$ is chosen to
be $20$. The initial stock and strike price are $x_0 = x_s =100$, and
the corresponding Black-Scholes price is ${\cal C}_0^{BS}=6.58$. The
number of basis functions is $M=8$. We run $500$ simulations
containing $500$ paths each, for which we extract the average price
and standard deviation on the price. An example of the result of
linear regression is plotted in Fig.~1. Each data point corresponds to
one trajectory of the Monte-Carlo at one instant of time $k$, and
represents the quantity:
\be\label{eqfig1}
e^{-\rho}{\cal C}_{k+1}(x_{k+1})+\phi_k(x_k)[x_k-e^{-\rho}x_{k+1}],
\ee as a
function of $x_k$. The full line represents the result of the
least-squared fit, form which we obtain ${\cal C}_{k}(x_{k})$. We show
in the inset the corresponding hedge $\phi_k$, that was constrained in
this case to be the $\Delta$ hedge.
\begin{figure}
\psfrag{axisx1}[ct][ct]{\small $x$}
\psfrag{axisy1}[cb][cb]{\small ${\cal C}_{10}(x)$}
\psfrag{axisx2}[ct][ct]{\small $x$}
\psfrag{axisy2}[cb][cb]{\small $\phi_{10}(x)$}
\centerline{\epsfig{file=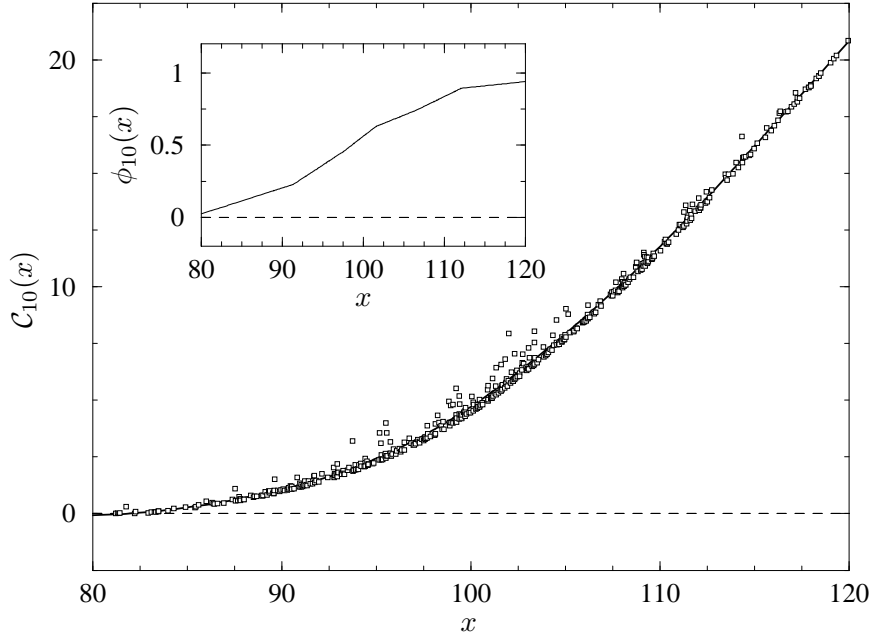,height=0.4\textheight}}
\caption{Option price as a function of underlying price for a {\sc
hmc} simulation. The full line corresponds to the fitted option price
${\cal C}_{10}(x)$ at the tenth hedging step ($k=10$) of a simulation
of length $N=20$. Square symbols correspond to the option price on the
next step, corrected by the hedge, Eq.~(\ref{eqfig1}) for individual
Monte-Carlo trajectories. Inset: Hedge as a function of underlying
price at the tenth step of the same simulation.}
\end{figure}

We obtain the following numerical results. For the {\sc rnmc}
(un-hedged) scheme, we obtain ${\cal C}_0^{RN}=6.68$ with a standard
deviation of $0.44$. For the {\sc hmc}, we obtain ${\cal
C}_0^{H}=6.55$ with a standard deviation of $0.06$, seven
times smaller than with the {\sc rnmc}. This variance reduction is
illustrated in Fig.~2, where we show the histogram of the MC results
both for the un-hedged case (full bars) and for the hedged case (dotted
bars).

Now we set the drift to $30\%$ annual. The Black-Scholes price,
obviously, is unchanged. A naive un-hedged Monte-Carlo scheme with the
objective probabilities would give a completely wrong price of
$10.72$, $60\%$ higher than the correct price, with a standard
deviation of $0.56$. On the other hand, the {\sc hmc} indeed produces
the correct price ($6.52$) with a standard deviation of $0.06$.  The
{\sc rnmc} scheme in this case simply amounts to setting `by hand' 
the drift to
risk-free rate, and therefore obviously gives back the above figures.
\begin{figure}
\psfrag{axisx1}[cb][cb]{\small ${\cal C}$}
\psfrag{axisy1}[ct][ct]{\small $N$}
\psfrag{long_long_long_long_legend1}{\small {\sc rnmc} price}
\psfrag{long_long_long_legend2}{\small {\sc hmc} price}
\psfrag{long_long_long_legend3}{\small Black-Scholes price}
\centerline{\epsfig{file=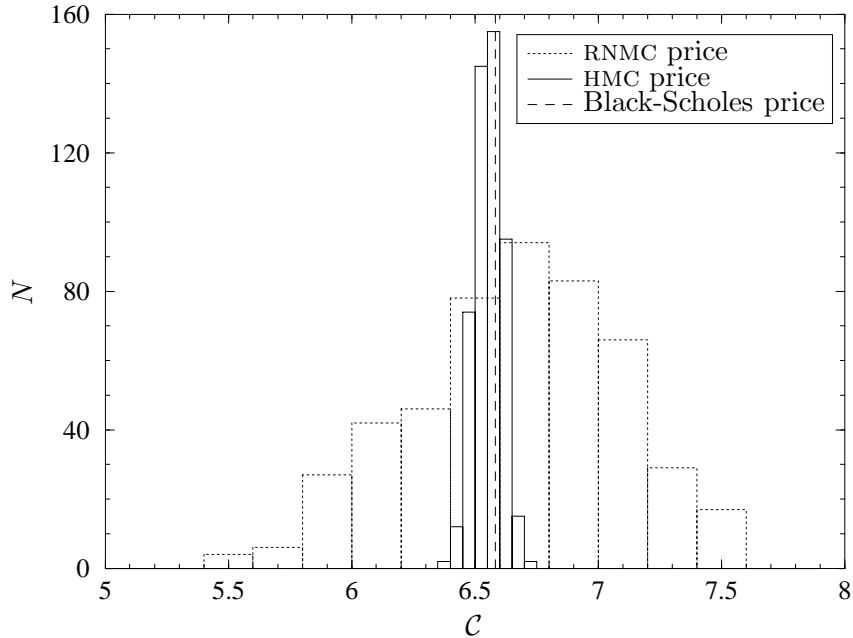,height=0.4\textheight}}
\caption{Histogram of the option price as obtained of 500 MC simulations
with different seeds. The dotted histogram corresponds to the {\sc rnmc} and
the full histogram to the {\sc hmc}. The dotted line indicates the exact
Black-Scholes price. Note that on average both methods give the correct price, but
that the {\sc hmc} has an error that is more than seven time smaller than
that of the {\sc rnmc}.}
\end{figure}

Therefore, we have checked that in the case of a geometric random
walk, the {\sc hmc} indeed gets rid of the drift and reproduces the
usual Black-Scholes results, as it should. This allows us to
confidently extend the method to other type of options and other
random processes.

\section{American and other exotic options}

The {\sc hmc} method can be used to reduce the Monte Carlo error for
all types of exotic options. We illustrate this point by showing how
the method can be extended to price an American put option.  In order to
implement the early exercise condition, one can simply replace ${\cal
C}_{k+1}(x_{k+1})$ in Eq.~(\ref{EQ4}) by $\max({\cal
C}_{k+1}(x_{k+1}),x_s-x_{k+1})$, where $x_s$ is the strike price. We
have chosen a slightly different implementation, where we first find
the early exercise point $x^*_{k+1}$ and exercise all options for
which $x_{k+1}<x^*_{k+1}$.

We have tested the method on a one year American put option on a stock
following a standard log-Brownian process. We follow the choice of
parameters made in \cite{Schwartz} in order to compare our results
with theirs.  The initial price and the strike are set to
$x_0=x_s=40$, the volatility to $20\%$ annual and the risk-free rate
and the drift to $6\%$.  As a benchmark price we use the value $2.314$
(quoted in \cite{Schwartz}) computed using a very accurate finite
difference method.  We computed the price within the {\sc hmc} using
$N_{\rm MC}=500$ paths and $M=8$ basis functions. To measure the
accuracy of the method, we ran the Monte Carlo 500 times with
different random seeds. The average price found was $2.302$ with a
standard deviation (around the true value $2.314$) of $0.032$.

We also used the least-square method ({\sc lsm}) of \cite{Schwartz}
with the same parameters ($N_{\rm MC}=500$ paths and $M=8$ basis
functions). Note that the {\sc lsm} used in \cite{Schwartz} differs
from the {\sc hmc} in the following way: (i) the hedge is not used in
the least-square fit Eq.~(\ref{EQ4}), (ii) on subsequent iterations
the option price is kept as the discounted pay-off on that particular
path (final pay-off or early exercise value) and (iii) the final
option price is given as the average pay-off of all paths. In this
framework the least-square fit only serves to find the early exercise
point. Therefore, for European options, the {\sc lsm} is identical to
the standard {\sc rnmc}.

The average price within the {\sc lsm} is found to be $2.423$ with a
standard deviation around the true value of $0.170$, five times larger
than for the above quoted $0.032$ for the {\sc hmc}. These numbers are
compatible with those found in \cite{Schwartz} where the error quoted
is $0.01$, i.e.\ $17$ times smaller but with $200$ times more paths
and $2.5$ times more intermediate points.

Obviously, the same variance reduction would hold for other exotic
paths, as those discussed in \cite{Schwartz}. We have not investigated
in depth the optimal values to be given to the parameters $M$ and
$N_{\rm MC}$, or the choice of the basis functions that minimize the
computation cost for a given accuracy. These are implementation issues
that are beyond the scope of the present paper.

\section{Purely historical option pricing}

We now turn to the idea of a purely historical {\sc hmc} pricing
scheme. We price a one month (21 business days) option on Microsoft
Corp., hedged daily, with zero interest rates. We used 2000 paths of
length 21 days, obtained form the time series of Microsoft during the
period May 1992 to May 2000. The initial price is always renormalized
to $100$. We use a set of $M=10$ basis functions, and keep with the
simple $\Delta$-hedge. From our numerically determined option prices,
we extract an implied Black-Scholes volatility by inverting the
Black-Scholes formula and plot it as a function of the strike, in
order to construct an implied volatility smile. The result is shown in
Fig.~3, together with the error bars, obtained from the residual risk
of the hedged options\footnote{The error on the price is given by the
residual risk divided by the square-root of the number of independant
paths. Since we over-sample the paths, we use $2000\times 2/21\approx
190$ as the effective number of paths.}. The residual risk itself,
divided by the call or the put option price (respectively for out of
the money and in the money call options), is given in the inset. We
find that the residual risk is around $42 \%$ of the option premium at
the money, and rapidly reaches $100 \%$ when one goes out of the
money.  These risk numbers are comparable to those obtained on other
options of similar maturity (see \cite{BP}), and are much larger than
the residual risk that one would get from discrete time hedging
effects in a Black-Scholes world.

\begin{figure}
\psfrag{axisx1}[ct][ct]{\small $x_s$}
\psfrag{axisy1}[cb][cb]{\small $\sigma_{\rm imp}(x_s)$}
\psfrag{axisx2}[ct][ct]{\small $x_s$}
\psfrag{axisy2}[cb][cb]{\small ${\cal R}(x_s)/{\cal C}(x_s)$}
\centerline{\epsfig{file=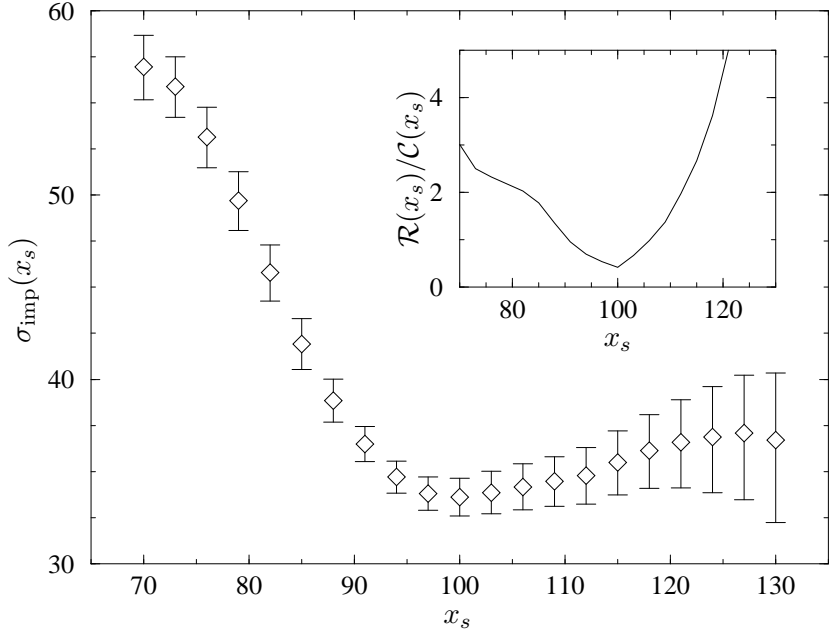,height=0.4\textheight}}
\caption{Smile curve for a purely historical {\sc hmc} of a one-month
option on Microsoft (volatility as a function of strike price).
The error bars are estimated from the residual Monte-Carlo error.
The inset shows the residual risk as a function of strike normalized by
the `time-value' of the option (i.e. by the call or put price, whichever
is out-of-the-money).}
\end{figure}
The smile that we obtain has a shape quite typical of those observed
on option markets. However, it should be emphasized that we have
neglected the possible dependence of the option price on the local
value of the volatility. This could be taken into account by letting
the function ${\cal C}_k$ depend not only on $x_k$ but also on the
value of some filtered past volatility $\sigma_k$.

\section{Conclusion and prospects}

We have presented what we believe to be a very useful Monte-Carlo
scheme, which closely follows the actual history of a trader hedged
portfolio. The inclusion of the optimal hedging strategy allows one to
reduce the financial risk associated to option trading, and for the
very same reason the variance of our {\sc hmc} scheme as compared to
the previously discussed {\sc rnmc} schemes. The explicit accounting
of the hedging cost naturally converts the objective probability into
the `risk-neutral' one.  This allows a consistent use of purely
historical time series to price derivatives and obtain their residual
risk. We believe that there are many extensions and applications of
the scheme, for example to price interest rate derivatives with
faithful historical models (such as the one proposed in
\cite{usrates}), and market hedging instruments. With some modifications and extra numerical cost, the
method presented here could be used to deal with transaction costs, or
with non quadratic risk measures (VaR hedging).

\subsection*{Acknowledgments}
We thank J.-P. Aguilar, A. Matacz and M. Bazant for interesting discussions.

\end{document}